\documentclass[prl,aps,superscriptaddress,footinbib,twocolumn]{revtex4}
\usepackage[latin9]{inputenc}
\usepackage{color}
\usepackage{amsmath}
\usepackage{amssymb}
\usepackage{stmaryrd}
\usepackage{graphicx}
\usepackage[normalem]{ulem}
\usepackage{bm}
\usepackage[unicode=true,bookmarks=true,bookmarksnumbered=false,bookmarksopen=false,breaklinks=false,pdfborder={0 0 1},backref=false,colorlinks=true]
 {hyperref}
\hypersetup{
 linkcolor=magenta, urlcolor=blue, citecolor=blue, pdfstartview={FitH},  unicode=true}

\makeatletter

\usepackage{amsfonts}
\usepackage{tabularx}
\usepackage{dcolumn}
\usepackage{bm}
\usepackage{graphicx}
\usepackage{epstopdf}
\usepackage{times}
\usepackage{lineno}
\hypersetup{urlcolor=blue}

\def\be{\begin{equation}}
\def\ee{\end{equation}}
\def\bea{\begin{eqnarray}}
\def\eea{\end{eqnarray}}
\def\bal{\begin{aligned}}
\def\eal{\end{aligned}}

\makeatother

\begin{document}
\nolinenumbers
\title{Thermalization of Quantum Many-Body Scars in Kinetically Constrained Systems}

\author{Jia-wei Wang}
\affiliation{Anhui Province Key Laboratory of Quantum Network, University of Science and Technology of China, Hefei, 230026, China}
\affiliation{Synergetic Innovation Center of Quantum Information and Quantum Physics, University of Science and Technology of China, Hefei, 230026, China}
\affiliation{ Hefei National Laboratory, University of Science and Technology of China, Hefei 230088, China}

\author{Xiang-Fa Zhou}\email{xfzhou@ustc.edu.cn}
\affiliation{Anhui Province Key Laboratory of Quantum Network, University of Science and Technology of China, Hefei, 230026, China}
\affiliation{Synergetic Innovation Center of Quantum Information and Quantum Physics, University of Science and Technology of China, Hefei, 230026, China}
\affiliation{ Hefei National Laboratory, University of Science and Technology of China, Hefei 230088, China}
\affiliation{Anhui Center for Fundamental Sciences in Theoretical Physics, University of Science and Technology of China}

\author{Guang-Can Guo}
\affiliation{Anhui Province Key Laboratory of Quantum Network, University of Science and Technology of China, Hefei, 230026, China}
\affiliation{Synergetic Innovation Center of Quantum Information and Quantum Physics, University of Science and Technology of China, Hefei, 230026, China}
\affiliation{ Hefei National Laboratory, University of Science and Technology of China, Hefei 230088, China}

\author{Zheng-Wei Zhou}\email{zwzhou@ustc.edu.cn}
\affiliation{Anhui Province Key Laboratory of Quantum Network, University of Science and Technology of China, Hefei, 230026, China}
\affiliation{Synergetic Innovation Center of Quantum Information and Quantum Physics, University of Science and Technology of China, Hefei, 230026, China}
\affiliation{ Hefei National Laboratory, University of Science and Technology of China, Hefei 230088, China}
\affiliation{Anhui Center for Fundamental Sciences in Theoretical Physics, University of Science and Technology of China}

\begin{abstract} 

The phenomenon of quantum many-body scars (QMBS) has been studied both theoretically and experimentally, due to its unusual violation of the eigenstate thermalization hypothesis (ETH). In this paper, we extend the ETH to a new description based on the grand canonical ensemble to depict the thermal properties of QMBS models.
For this purpose, we embed the dynamics of kinetically constrained systems within the Lindblad-like master equation, and demonstrate that the violation of the ETH by scar eigenstates is related to their slow decay in the corresponding dissipative process. 
Within this open system description, we reformulate the ETH to demonstrate that both scar eigenstates and thermal ones exhibit thermalization governed by grand canonical statistics. Consequently, our revised ETH unifies scars and thermal states under a cohesive thermodynamic rule. Our work resolves the fundamental tension between constraint-induced non-ergodicity and thermalization paradigms, establishing a unified route to generalized thermalization for quantum many-body systems.
\end{abstract}

\maketitle


\section*{I.~~~~Introduction}
The eigenstate thermalization hypothesis (ETH)~\cite{eth1_1991,eth2_1994}, first proposed in the 1990s, provides a foundational perspective on thermalization in quantum systems. ETH posits that for a many-body Hamiltonian $H$, all eigenstates exhibit thermal behavior. That is,
the local characteristics of eigenstates should coincide with the microcanonical average at corresponding energies\cite{eth6_1996,eth7_1999}.
While ETH has been validated across numerous many-body systems\cite{eth8_2008,eth10_2009,eth11_2010,eth12_2013,eth13_2014}, remarkable exceptions exist, including strong violations in integrable systems\cite{Integrable1,Integrable2,Integrable3} and many-body localized (MBL) phase\cite{MBL1,MBL2,MBLscibull1,MBLscibull2}, as well as weak violations in systems exhibiting Hilbert space fragmentation (HSF)\cite{scarforall} or quantum many-body scars (QMBS)~\cite{quantumscarexp_2017,onsagerscar,scartower,fractionpxp,bosehubbard}.
The last case is particularly intriguing---while most eigenstates thermalize, a non-thermal subspace of scar eigenstates persists, scaling extensively with system size.
QMBS further separate into two distinct groups: (i) exactly solvable ones that constitute a Krylov subspace with well-defined algebraic structures~\cite{AKLT1,AKLT2,SGA1,scarforall,creationoperators1,creationoperators2,fractionpxp,breakexactscar}, and (ii) non-exactly-solvable scar models, which primarily include kinetically constrained systems, such as the well-known PXP model~\cite{quantumscarexp_2017, highspinPXP,spin1kitaevmodel,QMBSsyzh,fermihubbardscar,QMBS1Dspin,systematic1d}. In class (ii), significant efforts have been made to elucidate the non-thermal characteristics of such scar eigenstates~\cite{scarprojection,cjturner1,cjturner2,pxpsga,zhaihui,scarprojection}. 
Notably, in \cite{Yinglei1,Yinglei2,Yinglei3}, non-hermitian dynamics has also been employed to depict the non-thermal behavior of QMBS.

Kinetically constrained QMBS models exhibit anomalous non-thermal characteristics despite their non-integrability. 
The absence of rigorous algebraic structures leaves the mechanism of ETH violation unresolved, posing a fundamental challenge: to what extent can such system thermalize, and what statistical principles govern the persistent non-ergodicity of scar eigenstates? 
In this paper, we demonstrate that both scar and thermal eigenstates exhibit thermalization governed by grand canonical ensemble.


We focus on the kinetically constrained systems supporting QMBS.
By constructing dissipation channels that effectively acts as the kinetic constraints, we manage to describe general constrained systems with Lindblad-like master equations.
Consequently, we observe that within a small energy interval, the scar eigenstates and the thermal ones exhibit divergent decay characteristics during dissipative process: scar states display lower decay rates that distinctly reflect their non-thermal behavior.
Therefore, we revise the ETH based on the open system dynamics, utilizing grand canonical ensemble with constraint-dependent thermodynamic quantities.
We numerically verify that the grand canonical ensemble captures the local properties of both the scar eigenstates and the thermal ones.
Our results suggest that the constrained QMBS models discussed here exhibit thermalization in accordance with the grand canonical ensemble.

\begin{figure}
	\centering
	\includegraphics[width=8.5cm]{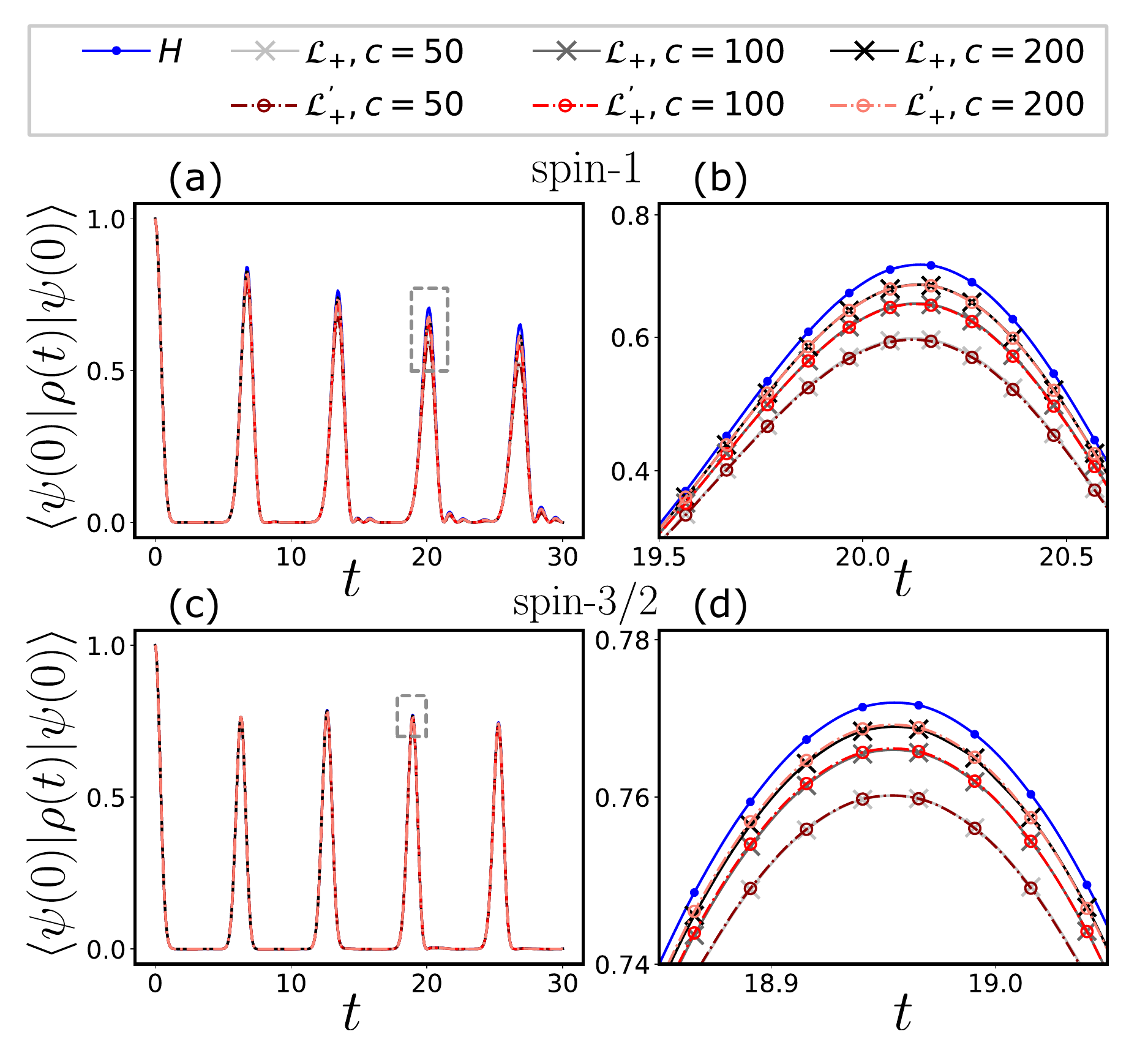}
	\caption{The evolution of the initial states' fidelity in the constrained system with Hamiltonian $H$, compared with that under the master equations marked by \(\mathcal{L}_{+}\) and \(\mathcal{L^{'}}_{+}\). In (a) and (b) we show the evolution of the model defined by Eqs.(\ref{Blockadingspin})-(\ref{DissipationQMBS}) in a spin-\(1\) chain with $9$ sites, while the plots (c) and (d) correspond to spin-$3/2$ chain with $7$ sites. We have chosen the initial state as $|\psi(0)\rangle$. Plots (b) and (d) present detailed views of the regions enclosed within the gray boxes in (a) and (c) respectively.}
	\label{fig: EvolvingtheInitialStates}
\end{figure}

\section*{II.~~~~Master equations governing general constrained systems}
A general constrained system can be described by a series of blockade $\mathcal{B}=\{\pi_1, \pi_2,\cdots,\pi_L\}$ acting on an unconstrained Hamiltonian $H_0=\sum_{k=1}^{N}h_k$, where $h_k$ is the $k$-th local Hamiltonian, and $\pi_m$'s are projectors satisfying $\pi_m^2=\pi_m$, $[\pi_m,\pi_n]=0$ for $\forall \pi_{m,n}\in \mathcal{B}$.Here $L$ and $N$ denote the total number of the projectors and local Hamiltonian $h_k$ respectively.
Then the projector onto the constrained Hilbert subspace ${\cal H}$ is $\hat{P}\equiv\prod_{k=1}^{L}(\mathbb{I}-\pi_k)$, and the constrained Hamiltonian writes:
\bea
  & H =\sum_{k=1}^{N}P_k h_k P_k=\sum_{k=1}^{N}(\mathbb{I}-\Pi_k)h_k(\mathbb{I}-\Pi_k) \label{BlockadingH} 
\\
  & \quad =\sum_{k=1}^{N}h_k-\Pi_k h_k - h_k \Pi_k + \Pi_k h_k \Pi_k,\label{Blockagingdetail}
\eea
where $P_k=\prod_{[\pi_m,h_k]\neq 0}(\mathbb{I}-\pi_m)$ denotes the constraint comprising all the blockades that do not commute with $h_k$, and $\Pi_k\equiv\mathbb{I}-P_k$ marks the local blockade acting on $h_k$.
$H$ commutes with $\hat{P}$ and serves as the Hamiltonian governing the constrained dynamics within ${\cal H}$, where its eigenstates satisfy $H|E_i\rangle=E_i|E_i\rangle,\ \hat{P}|E_i\rangle=|E_i\rangle$.
In Eq.(\ref{Blockagingdetail}), $\sum_{k=1}^{N}h_k$ remains the unconstrained Hamiltonian, and the rest terms constitute the blockade interactions. 
In particular, the term $\Pi_k h_k$ cancels the mapping from the constrained subspace ${\cal H}$ to the blockaded states, while $(\mathbb{I}-\Pi_k)h_k \Pi_k$ stops the latter mapping back to ${\cal H}$.
The back mapping terms vanish when acting on any states in ${\cal H}$ from left side, since we have $\Pi_k\hat{P}=(\mathbb{I}-P_k)\hat{P}=0$.
Hence these terms can be altered without affecting the dynamics within ${\cal H}$.
Specifically, we replace the term $(\mathbb{I}-\Pi_k)h_k \Pi_k$ with $\Pi_k h_k$'s anti-Hermitian conjugate and another imaginary part, and build up a non-Hermitian Hamiltonian as:
\be\label{Hn}
\bal
  H_N & =\sum_{k=1}^{N}h_k-\Pi_k h_k + h_k \Pi_k -i c \Pi_k \\
      & =H_0-\frac{i}{2}\sum_{\sigma,k=1}^{N}\gamma_{\sigma}L_{k,\sigma}^{\dagger}L_{k,\sigma},
\eal
\ee
here $i$ stands for the imaginary unit, and $c>0$ is a tunable real parameter. The terms $-\Pi_k h_k + h_k \Pi_k -i c \Pi_k$ align with the outcomes of the dissipative dynamics defined by two types of engineered dissipation channels $\sigma=\{1,2\}$ as:
\be\label{DissipationDefine}
\left\{\begin{array}{ll}
         \gamma_1=2c, & L_{k,1}=\Pi_k-\frac{i}{c}\Pi_k h_k \\
         \gamma_2=-\frac{2}{c},\ & L_{k,2}= \Pi_k h_k \\
       \end{array}\right.
\ee
$\gamma_{\sigma}$'s denote the dissipation rates, and $L_{k,\sigma}$'s are the dissspation operators.
Then we write out the Lindblad-like master equation for the dissipative system as:
\be\label{MEb}
  \bal
  \partial_t \rho & = \mathcal{L}(\rho) \equiv -i H_N \rho +i\rho H_N^{\dagger} +\mathcal{J}(\rho), \\
  \eal
\ee
where $\mathcal{L}$ is the Liouvillian superoperator acting on the system's density matrix (DM) $\rho$. The Hamiltonian $H_N$ shares the same right eigenvector $|E_i\rangle$ with $H$, and the jumping term $\mathcal{J}(\rho)=\sum_{k,\sigma} \gamma_{\sigma} L_{k,\sigma} \rho L_{k,\sigma}^{\dagger}=\sum_{k=1}^{N}2c\Pi_k \rho \Pi_k -2i\Pi_k h_k\rho \Pi_k +2i\Pi_k\rho h_k \Pi_k$ annihilates any $\rho$ within the constrained subspace. In principle, the two types of engineered dissipations combine perfectly to form the effective constraints, while within ${\cal H}$, the corresponding jumping terms of both types adequately cancel each other. Hence Eq.(\ref{MEb}) exhibits the same constrained dynamics as $H$, with ${\cal H}$ serving as a jump-free subspace (JFS). 
This JFS differs from the conventional decoherence-free subspace (DFS), which is annihilated by all dissipators simultaneously\cite{DFS1,DFS2}. 

\begin{figure}
	\centering
	\includegraphics[width=8.5cm]{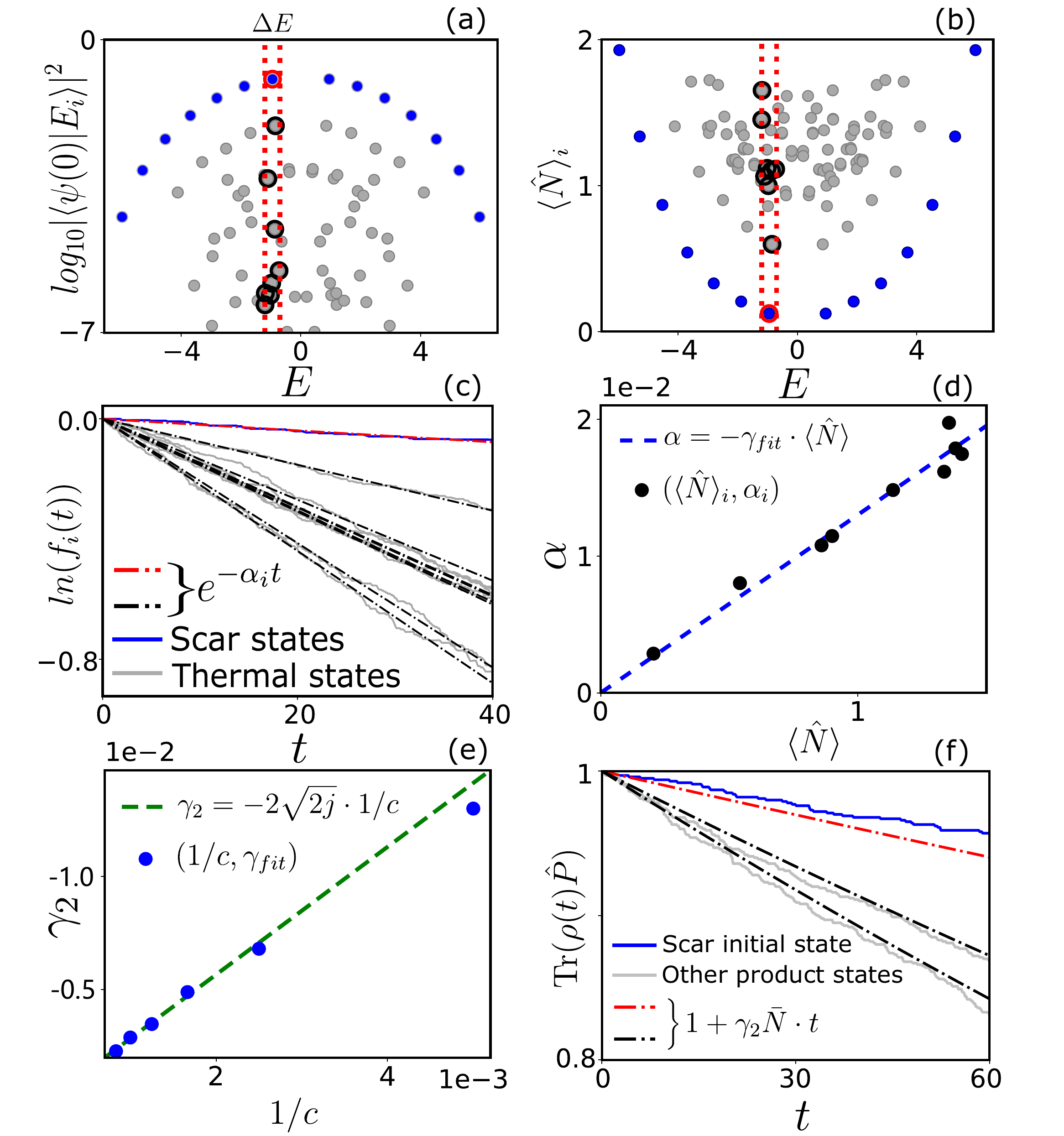}
	\caption{The exponential decay of the eigenstates' fidelity. Plots (a) and (b) show the eigenstates' overlaps with $|\psi(0)\rangle$, and their expectation values of $\hat{N}$. We pick $9$ eigenstates from a small energy interval $\Delta E$ and illustrate the evolutions of $f_i(t)$ in plot (c). The blue dots and the blue curve denote scar states, while the gray ones denote thermal states. The dashed lines are the fitted exponential functions, yielding a parameter $ \alpha_i $ for each $ |E_i\rangle $. In plot (d), for $9$ randomly picked eigenstates, we perform a linear fit of $ \alpha_i $ against $ \langle \hat{N} \rangle_i $, obtaining a near-linear relation: $ \alpha_i\approx-\gamma_{fit}\cdot \langle \hat{N} \rangle_i $. Plots (c) and (d) are computed with the parameter $ c = 200 $. In plot (e), by increasing $ c $ to $ 4, 6, 8, 10, 12 \times 10^2 $, $ \gamma_{fit} $ shows a near-linear relation with $ 1/c $: $\gamma_{fit}\approx\gamma_2=-2\sqrt{2j}\cdot 1/c $. Plot (f) shows the time evolution of $\text{Tr}(\rho(t)\hat{P})$ under $\partial_t \rho=\mathcal{L}'_+(\rho)$, with $ c=1200 $. The blue line indicates the initial state $|\psi(0)\rangle$, and the gray ones mark two other product states. Dashed lines indicate the estimated leakage $1+\gamma_2 \bar{N}t$. The calculations are executed on a spin-$1$ chain with $7$ sites, employing $500$ quantum trajectories~\cite{QTD,QTD1,QTD2,QTD3}.}
	\label{fig: EigenDecay}
\end{figure}

In contrast to the Lindblad master equation, where the dissipation rates are small and positive according to the Born-Markov approximation, the coefficient $\gamma_1$ and $\gamma_2$ in Eq.(\ref{MEb}) have opposite signs, indicating the presence of information feedback and non-Markovian characteristics~\cite{nonMarkovJump}.
Notably, the coefficient $\gamma_2$ tends to zero as the parameter $c\rightarrow +\infty$. In this limit, we can employ only the positive dissipations, yielding a master equation $\partial_t\rho =\mathcal{L}_+(\rho)$ comprised of:
\bea\label{MEqmbsPositive}
\bal
 & H_{+}     \equiv H_0-\frac{i}{2}\sum_{k=1}^{N}\gamma_1 L_{k,1}^{\dagger} L_{k,1} =H_N -\frac{i}{c}\sum_{k=1}^{N}h_k\Pi_k h_k,\\
 & \mathcal{J_+}(\rho)    \equiv \sum_{k=1}^{N} \gamma_1 L_{k,1}\rho L_{k,1}^{\dagger} = \mathcal{J}(\rho)+\frac{2}{c}\sum_{k=1}^{N}L_{k,2} \rho L_{k,2}^{\dagger}.
\eal
\eea
In Eq.(\ref{MEqmbsPositive}), we express the jumping terms as the sum of two parts.
Physically, when the parameter $c$ is large, the quantum jumps $2c \Pi_{k}\rho \Pi_{k}$ from $\mathcal{J} (\rho)$ will confine the system to the constrained subspace ${\cal H}$ due to Zeno effect~\cite{Zeno1,Zeno2}. Meanwhile, the leakage out of ${\cal H}$ caused by the second term $2/c \sum_{k=1}^N L_{k,2} \rho L_{k,2}^{\dagger}=2/c \sum_{k=1}^N\Pi_k h_k \rho h_k \Pi_k$ is also relatively small.
Hence the dynamics governed by $\mathcal{L}_+ $ increasingly converge towards the constrained dynamics within ${\cal H}$.
Experimentally, $\mathcal{L}_+$ can be approximated by compensating the total constrained Hamiltonian $\hat{P}H\hat{P}$ with dissipation channels $\{\gamma '=-\gamma_2,L'_k=L_{k,2}\}$. 
Provided that the dissipation rate $\gamma '\propto 1/c$ is sufficiently small, this results in a Lindblad master equation $\partial_t\rho=\mathcal{L}'_+(\rho)$ satisfying the BMA~\cite{footnote1}.

In order to demonstrate the feasibility of our framework, we take a 1D spin model hosting QMBS as an example.
The unconstrained Hamiltonian comprises $N$ spins of size $j\geq 1$: $H_0=\sum_{k=1}^{N}s_k^x$. $s_k^x$ is the $k$-th spin operator in the $x$-direction.
The blockade is given by:
\be\label{Blockadingspin}
  \mathcal{B}=\{\pi_k = |x \rangle \langle x|_{k,k+1};\ k=1,2,\cdots N \},
\ee
$\pi_k$ denotes a projector of the product state $|x\rangle_{k,k+1}=|j\rangle_k\otimes |-j\rangle_{k+1}$ on $k$-th and $k+1$-th sites, with $|m\rangle_k$ the $z$-direction spin eigenvector at the $k$-th site: $s_k^z|m\rangle_k=m|m\rangle_k$, $m=-j,-j+1,\cdots,j$.
$\mathcal{B}$ prohibits the neighboring spins from occupying $|x\rangle_{k,k+1}$, and arranges the spins in a 1D chain with periodical boundary.
Based on our framework, the non-Hermitian Hamiltonian writes: $H_N=\sum_{k=1}^{N}s_k^x -\sqrt{j}M_k +\sqrt{j}M_k^{\dagger} -i c \sqrt{j/2}\pi_k$, where $M_k\equiv \frac{1}{\sqrt{j}}\pi_k(s_k^x+s_{k+1}^x) = |x\rangle\langle y|_{k,k+1}$ denotes the mapping from ${\cal H}$ to the blockaded region, with $|y\rangle_{k,k+1}=\frac{1}{\sqrt{2}}(|j\rangle_k\otimes|-j+1\rangle_{k+1}+|j-1\rangle_k\otimes|-j\rangle_{k+1})$.
Hence the dissipation channels of the corresponding Lindblad-like master equations can be defined as
\be\label{DissipationQMBS}
\left\{
\bal
  &\gamma_1=c \sqrt{2j}, & L_{k,1}=\pi_{k}-\frac{i\sqrt{2}}{c}M_{k}, \\
  &\gamma_2=-\sqrt{2j}\frac{2}{c}, & L_{k,2}=M_{k}.
\eal
\right.
\ee

Notably, we have made minor model-dependent adjustments regarding the spin model above, resulting in slight differences between the dissipations in Eq.(\ref{DissipationDefine}) and Eq.(\ref{DissipationQMBS}); detailed analysis is available in the supplementary materials.

This constrained spin model is known to host QMBS within ${\cal H}$, and displays a quasi-periodical revival of the QMBS initial state: $|\psi(0)\rangle=|j,j,\cdots,j\rangle$~\cite{QMBS1Dspin}.
Besides, when the spin size is $j=1$, this model becomes equivalent to the PXP model~\cite{scarprojection,QMBS1Dspin}.
In Fig.~\ref{fig: EvolvingtheInitialStates}, we show the differences among $H$, $\mathcal{L}_+$ and $\mathcal{L}'_+$ via the evolution of the fidelity $\langle \psi(0)|\rho(t)|\psi(0)\rangle$. We confirm that as $c$ increases, the evolutions of the two master equations nearly overlap, and they gradually approach the unitary dynamics given by $H$.

\begin{figure*}
  \centering
  \includegraphics[width=14cm]{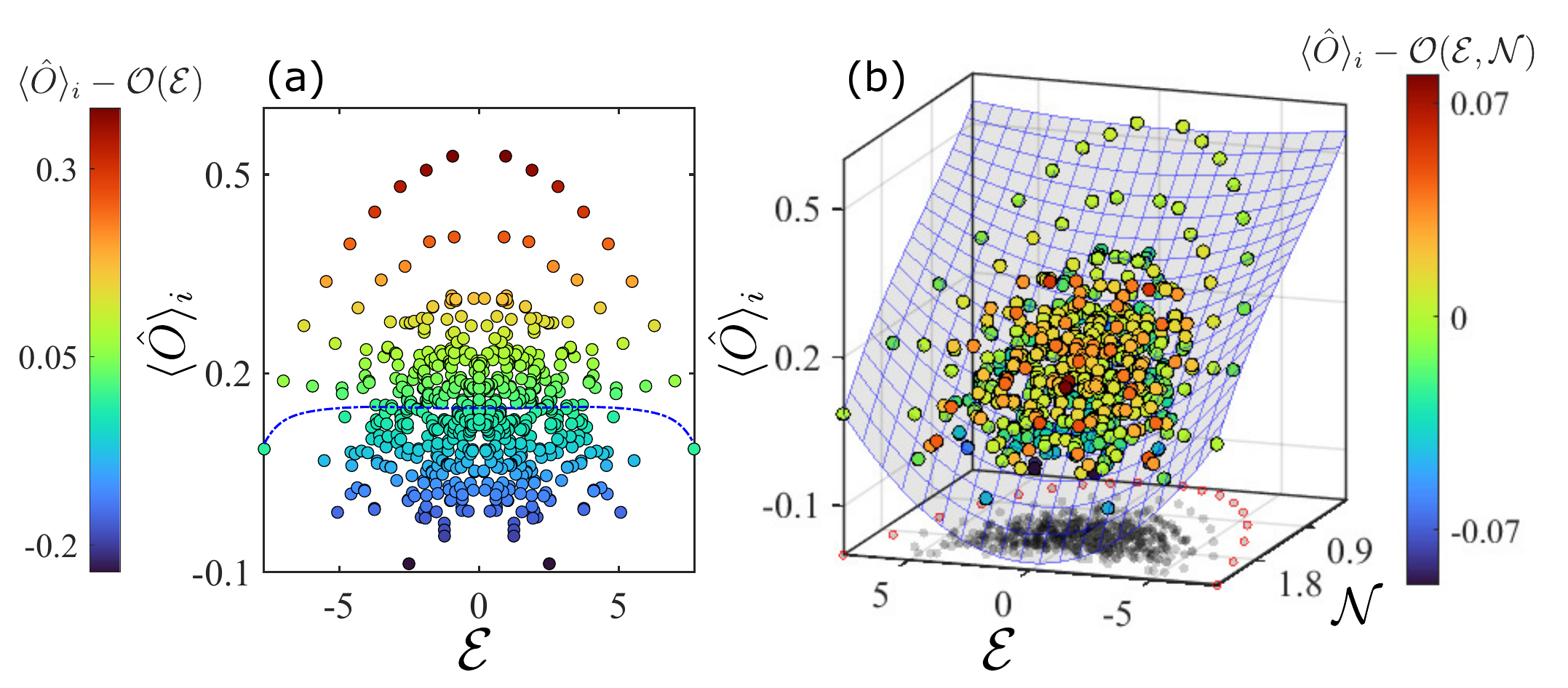}
  \caption{The expectation values $\langle \hat{O}\rangle_i$ for local observable $\hat{O}=s_k^z\otimes s_{k+1}^z$ in a spin-$1$ chain with $9$ spins. Plot (a) shows $\langle \hat{O}\rangle_i$'s relation with the energy $\mathcal{E}$, with the blue curve being the canonical average $\mathcal{O}({\mathcal{E}})$. In plot (b), $\langle \hat{O}\rangle_i$'s are fitted by a bivariate cubic function to form a surface $\mathcal{O}(\mathcal{E},\mathcal{N})$ defined by $\mathcal{E}$ and quasi-particle number $\mathcal{N}$. The color bars show the deviations of $\langle\hat{O}\rangle_i$ from the curve (or surface): $\langle\hat{O}\rangle_i-\mathcal{O}(\mathcal{E})$ (or $\langle\hat{O}\rangle_i-\mathcal{O}(\mathcal{E},\mathcal{N})$) through a range of changing colors. In plot (b) we also demonstrate the energy-quasiparticle relation by projecting the points onto the $\mathcal{E}-\mathcal{N}$ plane, with scar states highlighted in red circles.}
 	\label{fig: Visualization}
\end{figure*}

\section*{III.~~~~Slow decay indicating violation of ETH}
The dissipative mechanism provides a novel perspective on the thermodynamics of constrained systems. Specifically, we find that the scar eigenstates appear to show exceptionally lower decay rates than the thermal ones during the dissipative process defined by $\partial_t \rho=\mathcal{L}_+(\rho)$.

Starting from an eigenstate $|E_i\rangle$ of $H$, the system is approximately fixed during the jumpless process, since $H_+$ is dominated by the Hamiltonian $H_N$.
Moreover, any quantum jump of $\mathcal{J}_+$ will project the system outside the constrained subspace, ensuring that the resultant state is located in the blockaded area. Hence the DM of the system approximately writes: $\rho(t)\approx f_i(t)|E_i\rangle \langle E_i| + (\mathbb{I}-\hat{P})\rho(t)(\mathbb{I}-\hat{P})$, with the fidelity $f_i(t)=\langle E_i |\rho(t) |E_i\rangle$ satisfying:
\be\label{decayrate}
\bal
 & \frac{d f_i(t)}{dt} = \langle \frac{d\rho}{dt} \rangle_i = -i\langle H_+ \rho -\rho H_+^{\dagger} \rangle_i \\
   & \hspace{.5cm} \approx \gamma_2 \sum_{k=1}^{N} \langle L_{k,2}^{\dagger}L_{k,2}\rangle_i \cdot f_i(t) =-\alpha_i f_i(t),
\eal
\ee
where $\langle\cdot\rangle_i=\langle E_i| \cdot |E_i\rangle$ denotes the expectation value under the eigenstate $|E_i\rangle$. We note that the jumping term $\mathcal{J}_+(\rho)$ vanishes when acting on the eigenstates.
By defining $N_k=L_{k,2}^{\dagger}L_{k,2}$ and $\hat{N}=\sum_{k=1}^{N}N_k$, we obtain $f_i(t)\approx e^{-\alpha_i t}$ with the decay rate $\alpha_i\approx -\gamma_2 \langle \hat{N}\rangle_i$. For the spin model introduced above, $N_k=M_k^{\dagger} M_k$ is a projector on the neighboring state $|y \rangle_{k,k+1}$. $\hat{N}=\sum_{k=1}^{N}N_k$ counts the number of pattern $|y\rangle_{k,k+1}$ throughout the spin chain.


For eigenstates within a small energy interval $\Delta E$, Fig.~\ref{fig: EigenDecay}(c) illustrates the evolution of the fidelity $f_i(t)$, which is fitted with a exponential function $e^{-\alpha_i t}$. Then we confirm the proportional relation $\alpha_i\approx-\gamma_{fit} \langle \hat{N}\rangle_i$, and verify that $\gamma_{fit}$ approximates $\gamma_2=-2\sqrt{2j}/c$ in Fig.~\ref{fig: EigenDecay}(d)(e).
In Fig.~\ref{fig: EigenDecay}(c), the scar eigenstate shows a lower decay rate than the thermal states close in energy. Correspondingly, in Fig.~\ref{fig: EigenDecay}(b), the scar states demonstrate lower expectation values $\langle\hat{N}\rangle_i$, confirming their slower decay under the master equation $\partial_t\rho=\mathcal{L}_+(\rho)$.

Experimentally, this dissipative mechanism can be achieved by building a Markovian open system governed by $\partial_t \rho=\mathcal{L}'_+(\rho)$. For any initial state $|\phi\rangle=\sum_{i=1}^{N} a_i |E_i\rangle$ located in ${\cal H}$, the system will leak out of ${\cal H}$, with the leakage rate approaching $-\gamma_2 \bar{N}$. $\bar{N}=\sum_{i=1}^{N}|a_i|^2\langle \hat{N} \rangle_i$ is the long-time average of $\hat{N}$ during the unitary evolution. In Fig.~\ref{fig: EigenDecay}(f), we set the initial states to be product states with the same energy $\langle\phi| H|\phi\rangle=0$, and confirm this by calculating the evolution of the expectation value $\text{Tr}(\rho(t)\hat{P})$.

\begin{figure*}
	\centering
	\includegraphics[width=17.5cm]{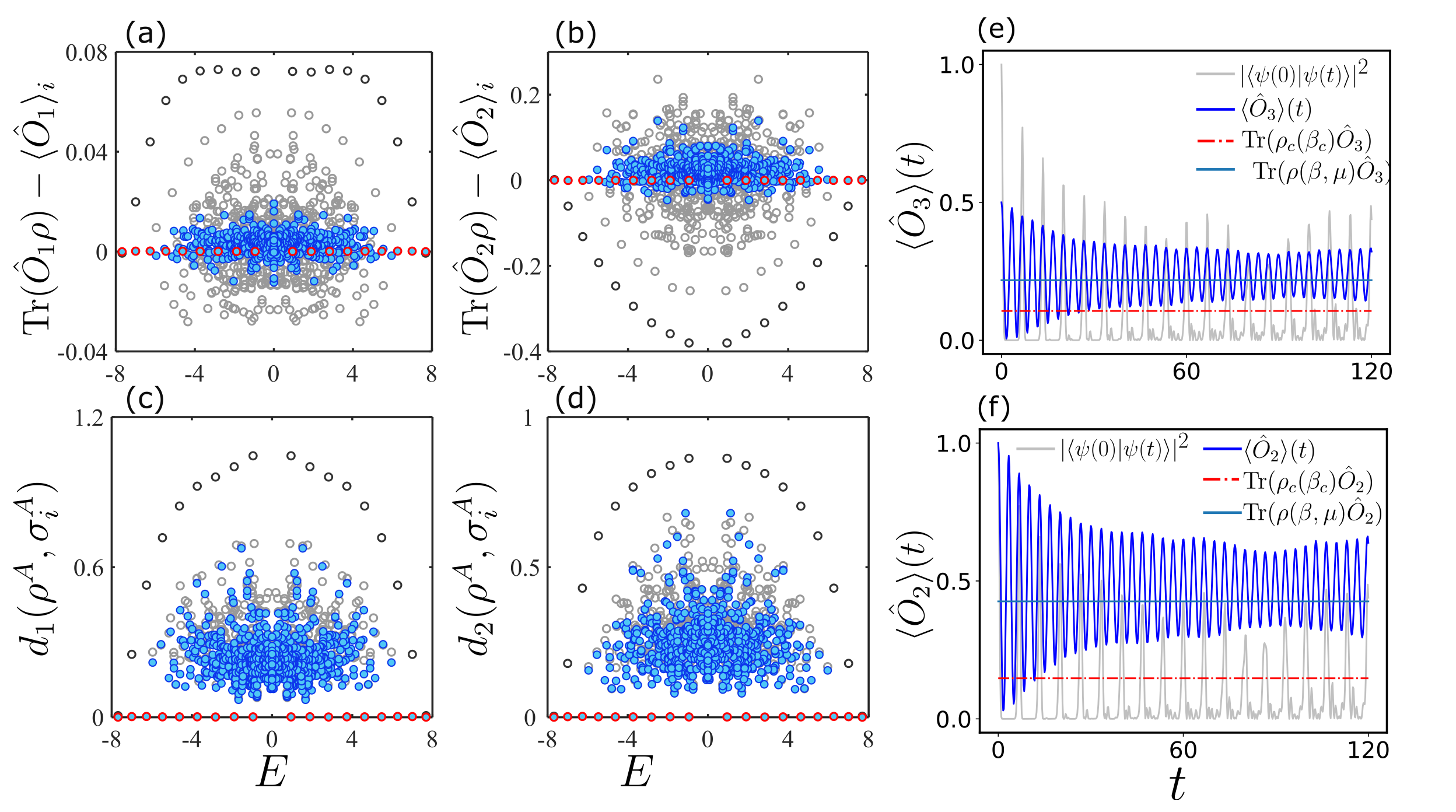}
	\caption{Deviations between the eigenstates and ensemble estimations. Plot (a) and (b) show the differences between the ensemble averages of local observables and their expectation values in the eigenstates \(|E_i\rangle\), expressed as \( \text{Tr}(\hat{O} \rho) - \langle \hat{O} \rangle_i \). Plot (c) and (d) present the Schatten distances \( d_p(\rho^A, \sigma_i^A) \) between the reduced DM \( \rho^A = \text{Tr}_{\bar{A}}(\rho) \) of ensembles and that of the eigenstates \( \sigma_i^A = \text{Tr}_{\bar{A}}(|E_i\rangle \langle E_i|) \) in the same \(2\)-sites subsystem $A$ for \( p = 1,2 \). In all plots (a)-(d), the horizontal axis corresponds to energy \( E \). Grey and black circles represent results from the canonical DM \( \rho = \rho_c(\beta_c) \), with darker circles indicating the scar states and lighter ones denoting the thermal states. Blue solid points represent results from the grand canonical DM \( \rho = \rho(\beta, \mu) \), with scar states highlighted in red circles. We have employed a spin-$1$ chain with 9 sites. Similar calculation in the spin-$3/2$ case can be found in supplementary materials. Plot (e) and (f) show the time evolution of expectation values for local observables. The initial state is chosen as \( |1,1,\cdots,1\rangle \). The grey solid lines show the quasi-periodical revival of the initial state, while the blue curves depict the time evolution of the local observables \( \hat{O}_3 \) and \( \hat{O}_2 \), expressed as \( \langle \hat{O}_{3,2} \rangle(t) = \langle \psi(t) | \hat{O}_{3,2} | \psi(t) \rangle \). The light blue straight lines indicate the averages of the observables derived from the grand canonical DM we constructed, and the red dashed lines represent the canonical average from ETH. We have employed the local observables \( \hat{O}_1 = |1\rangle \langle 1|_k + |0\rangle \langle 0|_k \) for plot (a), \( \hat{O}_2 = s_k^z \otimes s_{k+1}^z \) for plots (b) and (f), \( \hat{O}_3 \) as the projector on the state \( (|1\rangle_1 \otimes |1\rangle_2 - |-1\rangle_1 \otimes |-1\rangle_2)/\sqrt{2} \) for plot (e).}
	\label{fig: ExpectationandMatrixDist}
\end{figure*}

\section*{IV.~~~~Grand canonical thermalization theory.}
ETH posits that for a non-integrable many-body Hamiltonian $H$, the matrix elements of any local observable $\hat{O}$ in the energy eigenbasis should satisfy\cite{eth3_2011,eth4_2016,eth5_2018}:
\be\label{FullETH}
  \langle E_i| \hat{O} |E_{i'} \rangle=\mathcal{O}(\mathcal{E})\delta_{i,i'}+\Omega(\mathcal{E})^{-1/2}f(\mathcal{E},\omega)R_{i,i'},
\ee
where $\mathcal{E}\equiv (E_i+E_{i'})/2$ and $\omega\equiv E_i-E_{i'}$ denote the energies average and difference of the two eigenstates, and $\Omega(\mathcal{E})$ is the density of states at the energy $\mathcal{E}$. $\mathcal{O}(\mathcal{E})$ and $f(\mathcal{E},\omega)$ are smooth functions of $\mathcal{E}$ and $\omega$. $\delta_{i,i'}$ is the Kronecker delta function, while $R_{i,i'}$'s are erratically varying $O(1)$ numbers.
$\mathcal{O}(\mathcal{E})$ approaches the canonical average in the thermodynamic limit:
\be\label{eth}
   \mathcal{O}(\mathcal{E})\sim \langle \hat{O} \rangle_{\beta_c} = \text{Tr}(e^{-\beta_c H}\hat{O})/\text{Tr}(e^{-\beta_c H}),
\ee
where $\rho_c(\beta_c)=e^{-\beta_c H}/\text{Tr}(e^{-\beta_c H})$ is the canonical DM, and $\beta_c$ is the inverse-temperature calculated by fixing the energy: $\mathcal{E}=\text{Tr}(\rho_c (\beta_c)H)$.
If the above conditions are met, ETH asserts that for any physically preparable initial state $|\phi\rangle$, the long-time average of $\hat{O}$ converges to $\mathcal{O}(\mathcal{E})$, with $\mathcal{E}=\langle \phi|H|\phi\rangle$.

Based on the above discussions, We realize that the eigenstate $|E_i\rangle$ of the constrained systems can be interpreted as a dynamic equilibrium state in open systems at a certain energy $\mathcal{E}=E_i$. Besides, since the dissipative mechanism contributes to the state occupancy of the dynamic equilibrium, we can regard $N_k$ as a local 'quasi-particle' operator, and $\hat{N}=\sum_{k=1}^{N}N_k$ as the quasi-particle counter. Evidently, the energy $E$ and the quasi-particle number jointly permit an effective description of the equilibrium state.
Thus, for any eigenstate $|E_i\rangle$ within ${\cal H}$, we propose an adjustment to the diagonal terms from Eq.(\ref{FullETH}) by including another variable $\mathcal{N}$:
\be\label{AdjustedDiagonalterms}
\langle E_i |\hat{O} |E_i \rangle \sim \mathcal{O}(\mathcal{E},\mathcal{N}),
\ee
where $\mathcal{N}=\langle \hat{N} \rangle_i$ is the expectation value of the quasi-particle counter $\hat{N}$ under $| E_i\rangle$, $\mathcal{O}(\mathcal{E},\mathcal{N})$ is a smooth function of $\mathcal{E}$ and $\mathcal{N}$.
For a general constrained model, Eq.(\ref{AdjustedDiagonalterms}) states that the energy eigenstates' expectation value of any local observable should approximate the value of a smooth function uniquely determined by the eigenstate's energy and its quasi-particle number $\langle \hat{N} \rangle_i$.
In Fig.~\ref{fig: Visualization}, we demonstrate that the expectation values of the scar states cannot be unified with the majority of other eigenstates in a smooth function $\mathcal{O}(\mathcal{E})$.
However, all the expectation values can be unified in a smooth two-variable function $\mathcal{O}(\mathcal{E},\mathcal{N})$, certifying the above discussions.
Thus, all the eigenstates can be viewed as thermalized, as their local properties can be effectively described by Eq.(\ref{AdjustedDiagonalterms}).

Formally, the value of $\mathcal{O}(\mathcal{E},\mathcal{N})$ should coincide with the ensemble average of the following grand canonical DM:
\be\label{adjusteth}
  \rho(\beta,\mu)\equiv\frac{\sum_{i} e^{-\beta (E_i-\mu \langle \hat{N}\rangle_i)}|E_i\rangle\langle E_i|}{\sum_{i} e^{-\beta (E_i-\mu \langle \hat{N}\rangle_i)}},
\ee
where $\beta$ and $\mu$ are two thermodynamic variables describing the equilibrium. Specifically, $\beta$ is the inverse-temperature, and $\mu$ can be regarded as the chemical potential describing the dynamic equilibrium of quasi-particle exchange.
This grand canonical ensemble results from the dissipative mechanism employed to describe the constrained dynamics.
Concerning an eigenstate $|E_i \rangle$, $\beta$ and $\mu$ can be determined by fixing the energy and the quasi-particle number: $E_i=\text{Tr}(\rho(\beta,\mu )H), \langle \hat{N}\rangle_i=\text{Tr}(\rho(\beta,\mu )\hat{N})$.
Then we can estimate $\langle \hat{O}\rangle_i$ by calculating: $\mathcal{O}(\mathcal{E},\mathcal{N})\sim \text{Tr}(\rho(\beta,\mu)\hat{O})$.

We stress that our approach of defining quasi-particles deviates from the traditional route, where the grand canonical ensembles are introduced by regarding the conserved charges as the quasi-particles\cite{GGE1,GGE2,NATs1,NATs2,NATs3}. In contrast, the operator $\hat{N}$ defined above is not conserved, as it does not commute with the Hamiltonian $[H,\hat{N}]\neq 0$.
Further, our calculation indicates that considering only the diagonal portion of $\hat{N}$ is adequate for providing better estimation of relevant physical quantities.

Fig.~\ref{fig: ExpectationandMatrixDist} (a) and (b) illustrate the deviations between the ensemble average of local observables $\text{Tr}(\rho\hat{O})$ and their actual values $\langle \hat{O}\rangle_i$ for both the canonical DM $\rho_c(\beta_c)$ and the grand canonical DM $\rho(\beta,\mu)$ provided above.
In Fig.~\ref{fig: ExpectationandMatrixDist} (c) and (d), we also compute the Schatten distance $d_p$ between the local reduced DMs of the ensembles $\rho^A=\text{Tr}_{\bar{A}}(\rho(\beta,\mu))$(or $\text{Tr}_{\bar{A}}(\rho_c(\beta_c))$ and that of the eigenstates $\sigma_i^A=\text{Tr}_{\bar{A}}(|E_i\rangle\langle E_i|)$, which is defined as~\cite{AsignT_Eigen,Schattenp} $d_p(\rho, \sigma)\equiv\left\|\frac{\rho}{\|\rho\|_p}-\frac{\sigma}{\|\sigma\|_p}\right\|_p$,
with the Schatten $p$-norm $\|A\|_p\equiv\operatorname{Tr}\left(|A|^p\right)^{1 / p}$.
These results indicate that the grand canonical DM $\rho(\beta,\mu)$ provides a valid characterization for the local properties of all the eigenstates.
Moreover, the grand canonical ensemble gives extremely accurate descriptions of the scar eigenstates, whereas the canonical ensemble shows large deviations.
In Fig.~\ref{fig: ExpectationandMatrixDist} (e) and (f), we also show the time evolution of relevant operators  $\hat{O}_{2,3}$ starting from the initial state $|\psi(0)\rangle$.
It can be seen that $\langle \hat{O}\rangle(t)$ deviates significantly from the canonical average, while the grand canonical average gives a valid estimation for the long-time average of $ \hat{O}$.
Here the thermal coefficient $\beta$ and $\mu$ can be determined from the total energy and the long time average of the quasi-particle number $\bar{N}$, which can be obtained experimentally according to the previous section.

\section*{V.~~~~Conclusion}

\noindent In summary, we establish a mapping between general constrained models and engineered dissipative dynamics. Our strategy describes the constrained subspace as a jump-free region, where the jumping terms induced by the two types of dissipations cancel each other out. 
In this context, we find that the scar eigenstates display markedly lower decay rates than the thermal states, which distinctly reflects their non-thermal characteristics. This inspires us to reformulate the ETH by introducing the quasi-particle number $\mathcal{N}$ to the hypothesis, leading to a grand canonical equilibrium. The revised ETH successfully predicts local observables for all eigenstates and render the QMBS model thermalized under the principle of grand canonical ensemble. However, whether such thermalization mechanism is applicable to exactly solvable QMBS and many-body localized phases remains open. In addition, although the non-integrable constrained models discussed here show evidence of grand canonical thermalization, integrability is typically associated with the suppression of thermalization processes. A fundamental question is whether integrability can also suppress this grand canonical thermalization. Our work provides a unified pathway for exploring thermalization phenomena in a broader context.

\vspace{.5cm}
\textbf{Conflict of interests} The authors declare no competing financial interests.

\vspace{.3cm}
\textbf{Acknowledgement} We would like to thank Lei Ying and Xing-Shuo Xu for helpful discussions, and Xi-Wang Luo, Pan Gao for their assistence with the calculation. This work was funded by National Natural Science Foundation of China (No. 12474366, Grants No. 11974334, and No. 11774332), and Innovation Program for Quantum Science and Technology (Grant No. 2021ZD0301200). X.F.Z. We also acknowledge support from CAS Project for Young Scientists in Basic Research (Grant No. YSBR-049).

\vspace{.3cm}
\textbf{Conflict of interests} The authors declare no competing financial interests.


\vspace{.3cm}
\textbf{Author contributions} Z.-W.Z. provide the basic inspiration. X.-F.Z. and J.-W.W. developed the theoretical framework. J.-W.W. conduct the numerical calculations. All authors contributed to the discussions of the results and development of the manuscript.

\end{document}


\nolinenumbers
\renewcommand{\bibnumfmt}[1]{[S#1]}
\renewcommand{\citenumfont}[1]{S#1}
 \setcounter{equation}{0}
 \setcounter{figure}{0}
        
    \renewcommand{\theequation}{S\arabic{equation}}
    \renewcommand{\thefigure}{S\arabic{figure}}
\title{Supplementary Materials: Thermalization of Quantum Many-Body Scars in Kinetically Constrained Systems}
\setcounter{figure}{0}
\setcounter{table}{0}
\renewcommand\thefigure{S\arabic{figure}}
\renewcommand\thetable{S\arabic{table}}

\maketitle

In this supplementary material, we provide some derivation details from the main text, including the design of the non-Hermitian Hamiltonian, the construction of dissipation operators and the mapping between the spin model and the PXP model. In addition, we present several numerical computations conducted in the context of spin-$3/2$ scenario, alongside investigations in another QMBS model.

\tableofcontents

\section{Details in the derivation of the master equation}

{\bf Proof of the equivalence between the master equation description and the original constrained Hamiltonian.} This equivalence has been demonstrated in section II. We provide a few details below, in order to clarify the deduction we have done above.

First, we show that the original Hamiltonian $H$ shares the same right eigenvector with the non-Hermitian Hamiltonian $H_N$. Since $H$ commutes with the projector $\hat{P}$ of the constrained subspace, $H$ can be diagonalized within the constrained subspace ${\cal H}$. Then we have:
\bea
H |E_i\rangle & =& H\hat{P}|E_i\rangle =  \sum_{k=1}^{N}\left(h_k-\Pi_k h_k- h_k \Pi_k +\Pi_k h_k \Pi_k \right ) \hat{P}|E_i\rangle \nonumber \\
&=&\sum_{k=1}^{N}\left  (h_k-\Pi_k h_k \right ) \hat{P}|E_i\rangle = \sum_{k=1}^{N}\left  (h_k-\Pi_k h_k +h_k \Pi_k -ic \Pi_k \right) \hat{P}|E_i\rangle \nonumber \\
& = & H_N |E_i\rangle, \label{equiHHndeduction}
\eea
here we have considered the fact that the local projector $\Pi_k$'s are orthogonal with $\hat{P}$: $\Pi_k\hat{P}=0$.

Secondly, we provide a detailed deduction of Eq.(5). We equip the master equation with Hamiltonian $H_0$ and the dissipation channels described in Eq.(4), then we get the non-hermitian effective Hamiltonian of the system as:
\bea
 H_n &\equiv & H_0 - i\sum_{k,\sigma} \frac{\gamma_{\sigma}}{2} L_{k,\sigma}^{\dagger}L_{k,\sigma} \nonumber \\
     & = & H_0 +\frac{i}{c} \sum_{k=1}^{N} h_k \Pi_k h_k - i c \sum_{k=1}^{N}(\Pi_k+\frac{i}{c} h_k\Pi_k)(\Pi_k-\frac{i}{c} \Pi_k h_k) \nonumber \\
     & = & \sum_{k=1}^{N} h_k -i c \Pi_k -\Pi_k h_k +h_k \Pi_k \nonumber \\
     & =& H_N  \label{Hndeduction}
\eea
with a tunnable parameter $c$.
The jumping term $\mathcal{J}(\rho)$ can be obtained in the same manner:
\be\label{Jumpdeduction}
\bal
    \mathcal{J}(\rho) =& \sum_{k,\sigma} \gamma_{\sigma} L_{k,\sigma} \rho L_{k,\sigma}^{\dagger} \\
  = & 2c\sum_{k=1}^{N} (\Pi_k-\frac{i}{c} \Pi_k h_k)\rho(\Pi_k+\frac{i}{c} h_k \Pi_k)  - \frac{2}{c} \sum_{k=1}^{N} \Pi_k h_k \rho h_k \Pi_k \\
 = & \sum_{k=1}^{N} 2c\Pi_k \rho \Pi_k -2i\Pi_k h_k \rho \Pi_k + 2i \Pi_k \rho h_k \Pi_k.
\eal
\ee

{\bf Dynamics with only the positive dissipation rates.} Having demonstrated the deduction of the master equation, we turn to the master equation with only the positive dissipation rates. 
Formally, when $c \rightarrow \infty$, we can ignore the term with negative dissipation rates.
The non-Hermiatian Hamiltonian writes:
\be\label{Hpositivededuction}
   H_+ \equiv H_0-\frac{i}{2}\sum_{k=1}^{N}\gamma_1 L_{k,1}^{\dagger} L_{k,1} =H_N-\frac{i}{c}\sum_{k=1}^{N} h_k \Pi_k h_k,
\ee
and the jumping term $\mathcal{J}_+(\rho)$ reads:
\be\label{Jpositivededuction}
   \mathcal{J}_+(\rho) \equiv \sum_{k=1}^{N} \gamma_1 L_{k,1}\rho L_{k,1}^{\dagger} = \mathcal{J}(\rho)+\frac{2}{c} \sum_{k=1}^{N}\Pi_k h_k \rho h_k \Pi_k,
\ee
where the states within ${\cal H}$ leak out through the term $\Pi_k h_k \rho h_k \Pi_k$ to the blockade region, and then suffer a fast decay caused by the quantum jumps $2c \Pi_{k}\rho \Pi_{k}$ in $\mathcal{J}(\rho)$. Still, the superoperator $\mathcal{L}_{+}$ does not compose a standard Lindblad master equation, since the dissipation rate $\gamma_1=2c$ is set to be a large parameter in the following discussions, which is not consistent with the standard Markovian approximation.
If we keep the leakage caused by the term $\Pi_k h_k \rho h_k \Pi_k$, and replace the confinement caused by the Zeno effect with the original blockade effect defined by $\hat{P}$, we will arrive at a qualified Lindblad master equation $\partial_t\rho=\mathcal{L}'_{+}(\rho)$, with the constrained Hamiltonian $H_B=\hat{P}H\hat{P}=\hat{P}H_0\hat{P}$ governing the unitary dynamics and dissipation channels:
\be\label{dissipationchannelLp2}
\{\gamma '=-\gamma_2=\frac{2}{c},\ L'_k=L_{k,2}=\Pi_k h_k \},
\ee
then the non-Hermitian Hamiltonian reads:
\be\label{Hpositive2deduction}
\bal
   H'_+ & \equiv H_B-\frac{i}{2}\sum_{k=1}^{N}\gamma ' L_k^{\prime \dagger} L'_k=H_B-\frac{i}{c}\sum_{k=1}^{N} h_k \Pi_k h_k,  \\
\eal
\ee
and the jumping term $\mathcal{J}'_+$ can be rewriten as:
\be\label{Jpositive2deduction}
\bal
   \mathcal{J}'_+(\rho)    & \equiv \sum_{k=1}^{N} \gamma ' L'_k\rho L_k^{\prime\dagger} = \frac{2}{c}\sum_{k=1}^{N} \Pi_k h_k \rho h_k \Pi_k.
\eal
\ee
The dissipation rate is inversely proportional to the parameter $c$. When $c$ is large, this dissipation parameter can be very small,  so that the evolution of the system approaches the unitary dynamics defined by $H_B$ in the limiting case of $c\rightarrow \infty$.

\begin{figure}
	\centering
	\includegraphics[width=10cm]{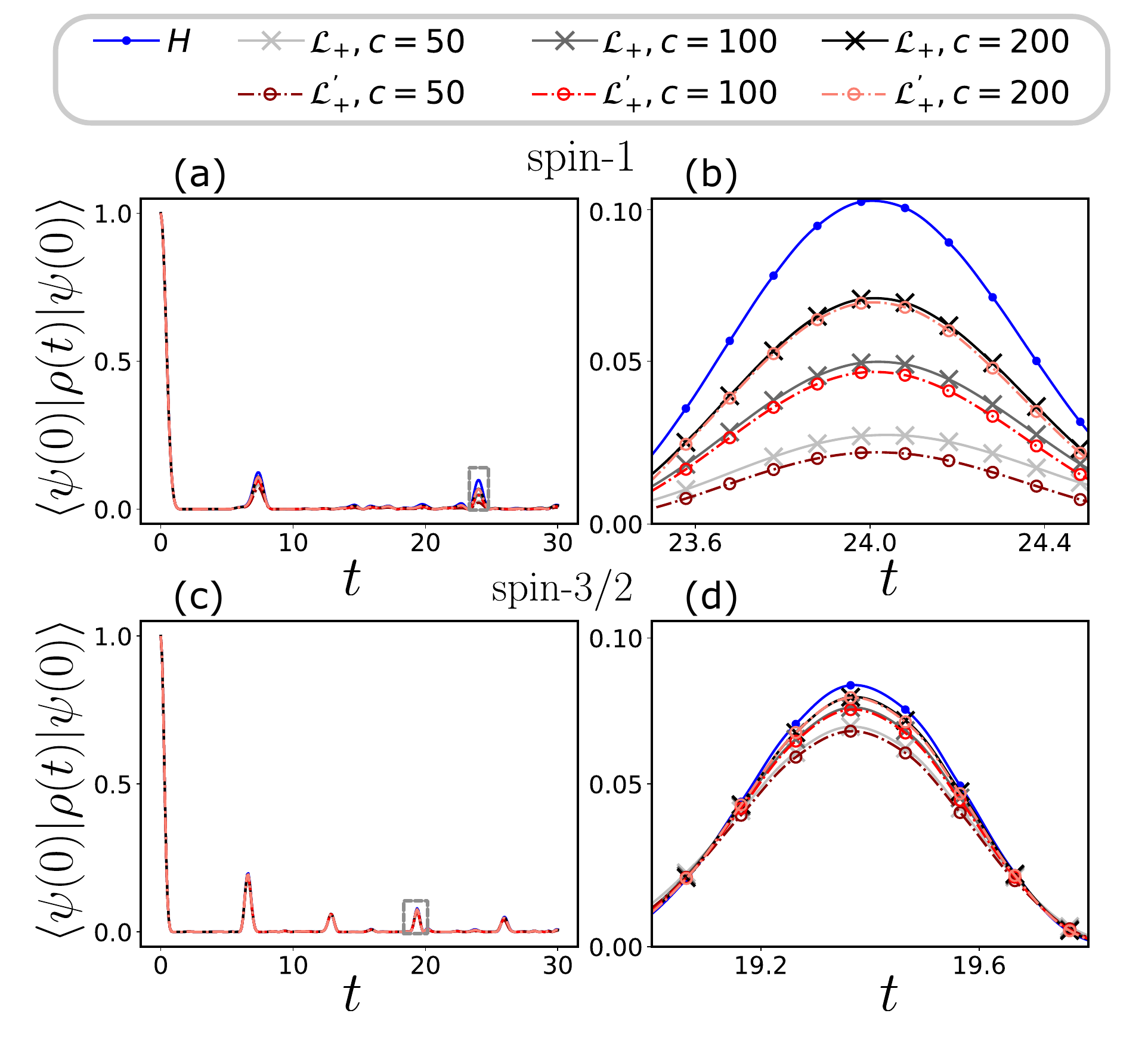}
	\caption{{\bf The evolution of the initial states' fidelity} in constrained system with Hamiltonian $H$, compared with that under the master equations marked by \(\mathcal{L}_{+}\) and \(\mathcal{L^{'}}_{+}\). In upper plots (a) and (b) we show the evolution in spin-\(1\) chain with $9$ sites, while the lower plots (c) and (d) correspond to spin-$3/2$ chain with $7$ spins. In the plots (a) and (c), we choose the initial states as random product states other than $|j,j,\cdots,j\rangle$. The left-side plots (b) and (d) present detailed views of the regions enclosed within the gray boxes in the corresponding figure on the right. }
	\label{fig: InitialStatesAPP}
\end{figure}

In Fig.~1 we have illustrated the differences among the dynamics of $H$, $\mathcal{L}_+$ and $\mathcal{L}'_+$, employing the spin model with the initial state $|\psi(0)\rangle=|j,j,\cdots,j\rangle$. 
Our calculations show that when c is large, both dynamical descriptions based on $\mathcal{L}_+$ and $\mathcal{L}'_+$ can approximate the ideal Hermitian dynamical evolution very well. 
Here we also confirm this result by providing the evolution of these dynamical equations under non-QMBS initial states, as shown in Fig.~\ref{fig: InitialStatesAPP}

{\bf QMBS in a 1D spin chain.} 
In this section, we will provide some calculation details regarding the spin model discussed in the main text. 
Based on the specific form of the Hamiltonian $H_0=\sum_{k=1}^{N}s_k^x$ and blockades $\mathcal{B}=\{\pi_k=| x\rangle\langle x|_{k,k+1};k=1,2,\cdots,N\}$, we follow the framework defined in the main text and write the constrained Hamiltonian as:
\be\label{HermitianHamiltonianQMBSApp}
\bal
  H & =\sum_{k=1}^{N}(\mathbb{I}-\Pi_k)s_k(\mathbb{I}-\Pi_k)=\sum_{k=1}^{N}s_k-\Pi_k s_k - s_k \Pi_k \\
    & =\sum_{k=1}^{N}s_k^x-  \sqrt{\frac{j}{2}}(|j\rangle\langle j|_{k-1}\otimes |-j\rangle\langle -j+1|_{k} +h.c.)-\sqrt{\frac{j}{2}}(|j\rangle\langle j-1|_{k}\otimes |-j\rangle\langle -j|_{k+1} +h.c.) ,
\eal
\ee
where $\Pi_k=\pi_{k-1}+\pi_{k}$ stands for the constriants felt by the local Hamiltonian $s_k^x$. We note that the terms $\Pi_k h_k \Pi_k$ vanish for this model.
Subsequently, we can obtain the explicit form of the dissipative channel by rigorously following Eq.(5), thereby arriving at the associated master equation. 
However, the obtained dissipation terms acts on three adjacent lattice sites, akin to $\Pi_k$. To simplify computations and obtain more localized dissipations, we reformulate the Hamiltonian as:
\be\label{HHQMBSAppRe}
  H =\sum_{k=1}^{N}s_k^x -\sqrt{j}(|x\rangle_{k,k+1} \langle y|_{k,k+1}+h.c.) =\sum_{k=1}^{N}s_k^x -\sqrt{j}M_k -\sqrt{j}M_K^{\dagger},
\ee
where $M_k \equiv \frac{1}{\sqrt{j}}\pi_k(s_k^x+s_{k+1}^x) = |x\rangle\langle y|_{k,k+1}$ is the mapping from ${\cal H}$ to the blockaded region, with $|y\rangle_{k,k+1}=\frac{1}{\sqrt{2}}(|j\rangle_k\otimes|-j+1\rangle_{k+1}+|j-1\rangle_k\otimes|-j\rangle_{k+1})$ the state mapped from the prohibited state, as defined in the manuscript. 
Following the same logic as established in the framework, we then write out the non-Hermitian Hamiltonian as:
\be\label{nHQMBSApp}
H_N=\sum_{k=1}^{N}s_k^x -\sqrt{j}M_k +\sqrt{j}M_K^{\dagger} -i c \sqrt{\frac{j}{2}}\pi_k.
\ee
From which we can build up a set of dissipation channels acting on only nearest-neighboring sites, as given in Eq.(8).

Next, we provide the deduction and the specific form of the master equation. The non-hermitian Hamiltonian writes:
\be\label{Hnspindeduction}
\bal
 H_N = & H_0 - i\sum_{k,\sigma} \frac{\gamma_{\sigma}}{2} L_{k,\sigma}^{\dagger}L_{k,\sigma} \\
     = & H_0 - i c \sqrt{\frac{j}{2}} \sum_{k=1}^{N}(\pi_k+\frac{i\sqrt{2}}{c} M_k^{\dagger})(\pi_k-\frac{i\sqrt{2}}{c} M_k)+\frac{i\sqrt{2j}}{c} \sum_{k=1}^{N} M_k^{\dagger} M_k \\
     = & \sum_{k=1}^{N} s_k^x -i c \sqrt{\frac{j}{2}}\pi_k -\sqrt{j} M_k +\sqrt{j} M_k^{\dagger} -i\frac{\sqrt{2j}}{c} M_k^{\dagger}M_k + i\frac{\sqrt{2j}}{c} M_k^{\dagger}M_k \\
     = & \sum_{k=1}^{N} s_k^x -i c \sqrt{\frac{j}{2}}\pi_k -\sqrt{j} M_k +\sqrt{j} M_k^{\dagger},
\eal
\ee
which shares the same right eigenvectors with the Hamiltonian from Eq.(\ref{HHQMBSAppRe}) in the constrained subspace. We have used the relation $\pi_k M_k=M_k$ in the above deduction. Finally, we have the jumping terms:
\be\label{Jumpspindeduction}
\bal
    & \sum_{k,\sigma} \gamma_{\sigma} L_{k,\sigma} \rho L_{k,\sigma}^{\dagger} \\
  = & c\sqrt{2j}\sum_{k=1}^{N} (\pi_k-\frac{i\sqrt{2}}{c} M_k)\rho(\pi_k+\frac{i\sqrt{2}}{c} M_k^{\dagger}) - \sqrt{2j}\frac{2}{c} \sum_{k=1}^{N} M_k \rho M_k^{\dagger} \\
  = & \sum_{k=1}^{N} c\sqrt{2j} \pi_k \rho \pi_k -2i\sqrt{j} M_k \rho \pi_k + 2i\sqrt{j} \pi_k \rho M_k^{\dagger} +\sqrt{2j}\frac{2}{c} M_k \rho M_k^{\dagger} - \sqrt{2j}\frac{2}{c} M_k \rho M_k^{\dagger} \\
  = & \sqrt{2j} \sum_{k=1}^{N} c \pi_k \rho \pi_k -i \sqrt{2} M_k \rho \pi_k + i \sqrt{2} \pi_k \rho M_k^{\dagger}.
\eal
\ee
We can see that any state within the constrained subspace remains unaffected by the jumping term since it is annihilated by the blockades $\pi_k$'s from either side: $\pi_k\hat{P}=\hat{P}\pi_k=0$.
Therefore, the above master equation equivalently describes the dynamics of the Hamiltonian from Eq.(\ref{HHQMBSAppRe}).

We note that in the framework provided in the main text, we build up the dissipation channels based on single local Hamiltonian part $h_k$ and the blockades $\pi_k$'s that does not commute with $h_k$. On the contrary, here we build the dissipation channels in the spin model based on single blockade $\pi_k$ and local Hamiltonian terms that effectively act on it. For example, we have utilized the operator $M_k\propto \pi_k(s_k^x+s_{k+1}^x)$ instead of $\Pi_ks_k^x=(\pi_{k-1}+\pi_{k})s_k^x$ to describe the connection between the constrained subspace and the blockaded states.
Although the construction method utilized in the framework is universal, the approach applied to spin chains may provide significant convenience for subsequent discussions. Nevertheless, the latter necessitates a case-by-case examination when extending to other models.

{\bf Proportional relation between the eigenstate decay rates and the expectation values of $\hat{N}$.} We provide a detailed deduction of Eq.(9), demonstrating the decay rate of $H$'s eigenstates under the dynamical equation $\partial_t \rho=\mathcal{L}_+(\rho)$:
\be\label{decayratededuction}
\bal
 \frac{d f_i(t)}{dt} & =  \frac{d}{dt} \langle E_i|\rho(t)| E_i \rangle \\
 & =\langle E_i|\frac{d}{dt}\rho(t)| E_i \rangle=\langle E_i|\mathcal{L}_+(\rho(t))| E_i \rangle=-i\langle E_i|H_+\rho(t)| E_i \rangle+i\langle E_i|\rho(t)H_+^{\dagger}| E_i \rangle \\
 & =\sum_{k=1}^N -i\langle E_i|(2h_k\Pi_k-\frac{i}{c}h_k\Pi_k h_k)\rho(t)| E_i \rangle +i\langle E_i|\rho(t)(2\Pi_k h_k+\frac{i}{c}h_k\Pi_k h_k)| E_i \rangle.
\eal
\ee
We note that the jumping term $\mathcal{J}_+(\rho)$ vanishes due to the action of $\langle E_i|\cdot |E_i \rangle$. Then we make the following approximation to simplify the calculation. As $c$ approaches infinity, the eigenstates of the constrained Hamiltonian $H$ can be regarded as approximately equivalent to the right-eigenvectors of the non-Hermitian Hamiltonian $H_+$:
\be\label{EigenApproxiamtion}
  H|E_i \rangle  = E_i|E_i \rangle = H_N|E_i \rangle =(H_+ +\frac{i}{c}\sum_{k=1}^{N} h_k \Pi_k h_k)|E_i \rangle  \sim H_+|E_i \rangle .
\ee
Consequently, for the trajectory without jumps, the system remains in its initial state $|E_i\rangle$. Moreover, any quantum jump of $\mathcal{L}_+$ projects the system outside the constrained subspace, thereby ensuring that the resultant state is orthogonal to the initial one. 
Therefore, we have $\rho(t)\approx f_i(t)\cdot|E_i\rangle\langle E_i|+(\mathbb{I}-\hat{P})\rho(t)(\mathbb{I}-\hat{P})$, where $\hat{P}\rho'\hat{P}$ denotes the part of the DM that has been mapped outside the subspace ${\cal H}$ by quantum jumps.
Then we arrive at the following equation:
\be\label{decayratededuction2}
 \frac{d f_i(t)}{dt} \approx -\frac{2}{c}\sum_{k=1}^N\langle h_k\Pi_k h_k \rangle_i \cdot f_i(t) = \gamma_2\sum_{k=1}^N\langle L_{k,2}^{\dagger}L_{k,2} \rangle_i \cdot f_i(t).
\ee
Finally, we obtain $f_i(t)\approx e^{-\alpha_i t}$ with the decay rate $\alpha_i=-\gamma_2 \langle \hat{N}\rangle_i$, where $\hat{N}=\sum_{k=1}^{N}N_k$ and $N_k=\sum_{k=1}^{N}L_{k,2}^{\dagger}L_{k,2}$ as defined in section III.

{\bf Proportional relation between the leakage rate and the long-time average of $\hat{N}$.}
For any physically preparable initial state $|\psi(0)\rangle=\sum_{i=1}^{N}a_i |E_i\rangle$, we define the possibility of the system remaining in the subspace ${\cal H}$ during the dissipative process governed by $\partial_t\rho=\mathcal{L}'_+(\rho)$: $h(t)=\text{Tr}(\rho(t)\hat{P})$. Similarly we have:
\be\label{leakrate1}
\bal
  \frac{d h(t)}{dt} & =  \text{Tr}(\hat{P}\frac{d}{dt}\rho(t)) \\
 & =-\frac{\gamma '}{2}\sum_{k=1}^{N}\text{Tr}(\hat{P}(L_k^{\prime\dagger} L'_k \rho + \rho L_k^{\prime\dagger} L'_k))=\frac{\gamma_2}{2}\text{Tr}(\hat{P}(\hat{N} \rho + \rho \hat{N})), 
\eal
\ee
then we utilize the former assumption that $\rho(t)$ is approximately governed by $H$ for the process without jumps, and will be mapped out of the subspace ${\cal H}$ for any possible jump: $\rho(t)\approx h(t)\cdot \rho_{\cal H}(t)+(\mathbb{I}-\hat{P})\rho(t)(\mathbb{I}-\hat{P})$.
$\rho_{\cal H}(t)$ denotes the DM within ${\cal H}$, and can be approximated by $e^{-iHt}\rho(0)e^{iHt}$. Following this, we have:
\be\label{leakrate2}
\bal
  \frac{d h(t)}{dt} & \approx \gamma_2\text{Tr}(\hat{N}e^{-iHt}\rho(0)e^{iHt}) h(t), \\
  h(t) & \approx 1+\gamma_2 \int_{0}^{t} dt' h(t') \text{Tr}(\hat{N}e^{-iHt'}\rho(0)e^{iHt'}),
\eal
\ee
then we can keep the terms up to the first order of $\gamma_2\propto 1/c$ and get: $h(t)\approx 1+\gamma_2 \int_{0}^{t}\text{Tr}(\hat{N}e^{-iHt'}\rho(0)e^{iHt'})dt'$. Hence for a long-time process, we arrive at a leaking rate $\alpha_{L}=\gamma_2 \bar{N}$ and an exponential leaking process denoted by $h(t)\approx 1+\alpha_{L} t$, with $\bar{N}=\sum_{i=1}^{N}|a_i|^2\langle \hat{N} \rangle_i $.

\section{Mapping the spin-1 scenario to PXP model}

The spin-$1$ scenario of the 1D spin chain model can be mapped to the PXP model. Here we outline the derivation of the mapping from the Hamiltonian in Eq.(\ref{HHQMBSAppRe}) to the standard PXP Hamiltonian.
First, we rewrite Eq.(\ref{HHQMBSAppRe}) in the matrix form for spin-$1$ scenario as:
\bea
 H  &= & \sum_{k=1}^{N}s_k^x-M_k-M_k^{\dagger} \nonumber \\
    &= & \frac{1}{\sqrt{2}}\sum_{k=1}^{N}\left(\begin{array}{ccc}
                             0 & 1 & 0 \\
                             1 & 0 & 1 \\
                             0 & 1 & 0
                           \end{array}\right)_k  -\left(\begin{array}{ccc}
                           0 & 1 & 0 \\
                           1 & 0 & 0 \\
                           0 & 0 & 0
                           \end{array} \right)_k \otimes \left( \begin{array}{ccc}
                           0 & 0 & 0 \\
                           0 & 0 & 0 \\
                           0 & 0 & 1
                           \end{array} \right)_{k+1}\nonumber 
                           -\left(\begin{array}{ccc}
                           1 & 0 & 0 \\
                           0 & 0 & 0 \\
                           0 & 0 & 0
                           \end{array} \right)_k \otimes \left( \begin{array}{ccc}
                           0 & 0 & 0 \\
                           0 & 0 & 1 \\
                           0 & 1 & 0
                           \end{array} \right)_{k+1}\nonumber \\
    &= & \frac{1}{\sqrt{2}}\sum_{k=1}^{N}\left(\begin{array}{ccc}
                           0 & 1 & 0 \\
                           1 & 0 & 0 \\
                           0 & 0 & 0
                           \end{array} \right)_k \otimes \left( \begin{array}{ccc}
                           1 & 0 & 0 \\
                           0 & 1 & 0 \\
                           0 & 0 & 0
                           \end{array} \right)_{k+1}\nonumber 
                           + \left(\begin{array}{ccc}
                           0 & 0 & 0 \\
                           0 & 1 & 0 \\
                           0 & 0 & 1
                           \end{array} \right)_k \otimes \left( \begin{array}{ccc}
                           0 & 0 & 0 \\
                           0 & 0 & 1 \\
                           0 & 1 & 0
                           \end{array} \right)_{k+1},       \label{Hspin1scenario}
\eea
where the matrix expansion is written under the local spin basis of $s_k^z$: $\{|m=1\rangle,|m=0\rangle,|m=-1\rangle,\}$. Then we map these basis to two composite spin-$1/2$ sites as :
\be\label{Basistrans}
\bal
  & \left| m=1 \right\rangle_k=\left| \downarrow, \uparrow \right\rangle_{2k-1,2k}, \\
  & \left| m=0 \right\rangle_k=\left| \downarrow, \downarrow \right\rangle_{2k-1,2k},\\
  & \left| m=-1 \right\rangle_k=\left| \uparrow, \downarrow \right\rangle_{2k-1,2k}.
\eal
\ee
Then we rewrite the Hamiltonian from Eq.(\ref{Hspin1scenario}) in the spin-$1/2$ basis:
\bea
 H &= & \frac{1}{\sqrt{2}}\sum_{k=1}^{N}P_{2k-1}^{\downarrow}\otimes \sigma_{2k}^x \otimes P_{2k+1}^{\downarrow} + P_{2k}^{\downarrow}\otimes \sigma_{2k+1}^x \otimes P_{2k+2}^{\downarrow} \nonumber \\
   &= & \frac{1}{\sqrt{2}}\sum_{l=1}^{2N}P_{l-1}^{\downarrow}\otimes \sigma_{l}^x \otimes P_{l+1}^{\downarrow},   \label{Hspin1spinhalf}
\eea
where $P_l^{\downarrow}=|\downarrow\rangle \langle \downarrow |_l$ is the projector of spin-down state $|\downarrow\rangle_l$ on the $l$-th site, $\sigma_l^x$ is the Pauli operator on the $x$ direction. The final expression in Eq.(\ref{Hspin1spinhalf}) is exactly the Hamiltonian of a PXP model with $2N$ sites.

\begin{figure}
	\centering
	\includegraphics[width=8.5cm]{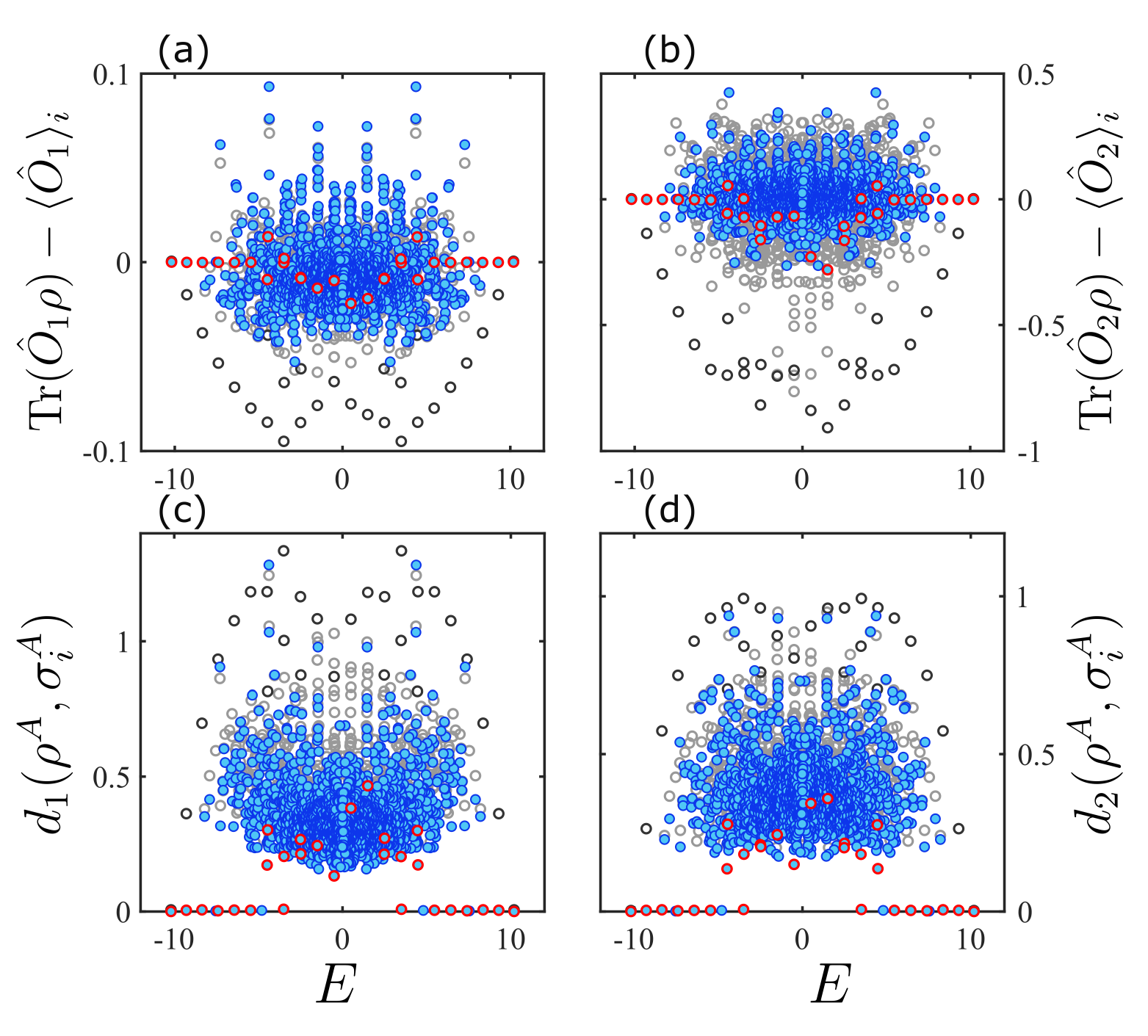}
	\caption{{\bf Deviations between the eigenstates and the ensemble estimations for spin-$3/2$ case.} Plot (a) and (b) show the differences between the DM average of local observables and their expectation values in the eigenstates \(|E_i\rangle\), expressed as \( \text{Tr}(\hat{O} \rho) - \langle \hat{O} \rangle_i \). In (a), the local observable is chosen as \( \hat{O}_1 = |3/2\rangle \langle 3/2|_k \). In (b), the local observable is also \( \hat{O}_2 = s_k^z \otimes s_{k+1}^z \). Plot (c) and (d) present the Schatten distances \( d_p(\rho^A, \sigma_i^A) \) between the reduced DM \( \rho^A = \text{Tr}_{\bar{A}}(\rho) \) and the reduced DM of the eigenstates \( \sigma_i^A = \text{Tr}_{\bar{A}}(|E_i\rangle \langle E_i|) \) in the same subsystem $A$, where we take \( p = 1,2 \). In all plots (a)-(d), the horizontal axis corresponds to energy \( E \), the grey and black circles represent results from the canonical DM \( \rho = \rho_c(\beta_c) \), with darker circles indicating scar states and lighter ones denoting thermal states. Blue solid points represent results calculated from the grand canonical DM \( \rho = \rho(\beta, \mu) \), with scar states highlighted in red circles. We have employed a spin-$3/2$ chain with 7 sites.}
	\label{fig: spin15expectation_and_schatten_distance}
\end{figure}

\begin{figure}
	\centering
	\includegraphics[width=8.5cm]{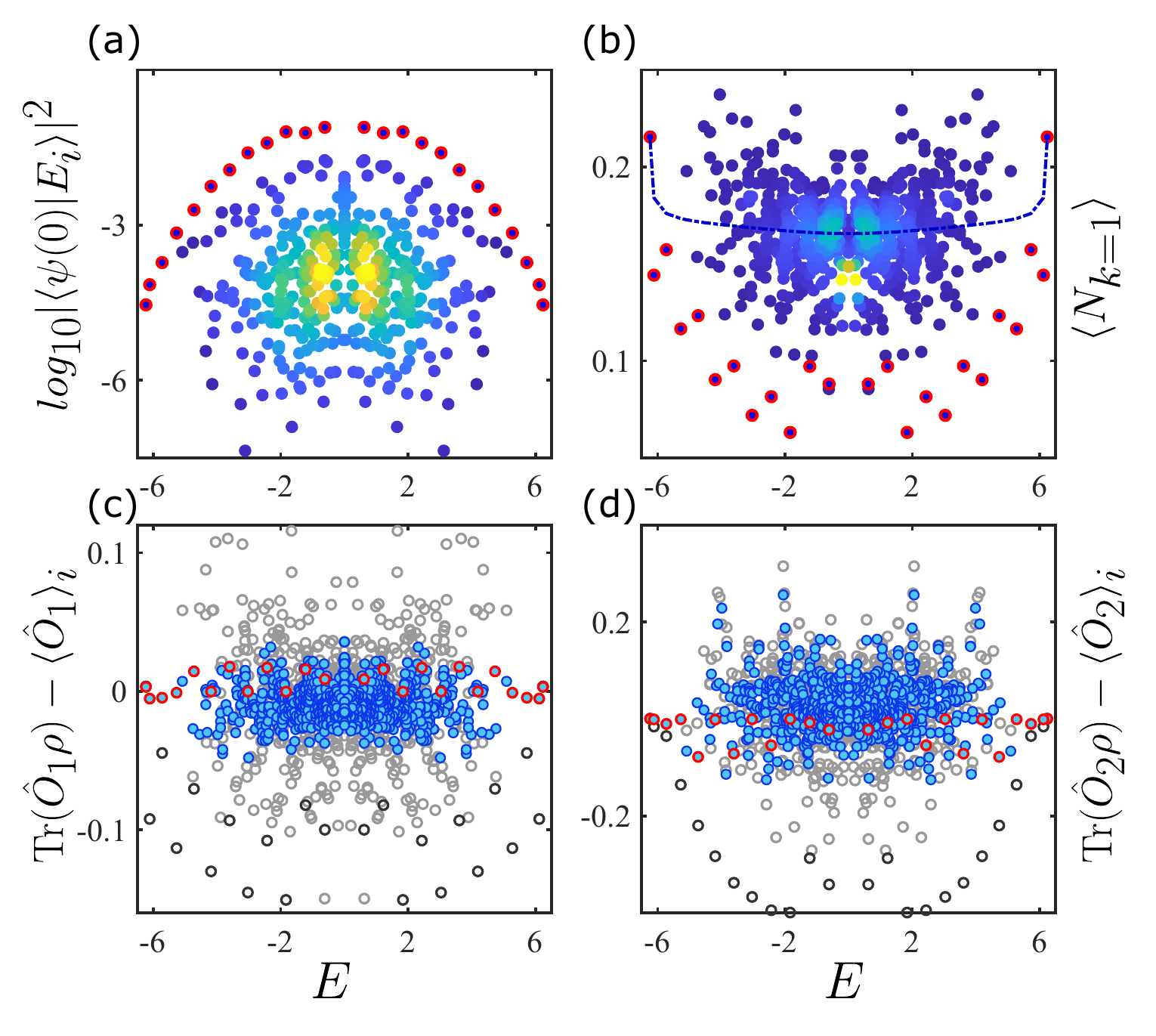}
	\caption{{\bf High-dimensional PXP model.} Plot (a) shows the overlap between the eigenstates and the initial state $|\psi(0)\rangle =|j,-j,\cdots,j,-j\rangle$. Plot (b) shows the expectation values of the local observable \(N_{k=1}\), with the points indicating the results obtained from the eigenstates: \(\langle N_{k=1} \rangle_i = \langle E_i | N_{k=1} | E_i \rangle\), and the dashed line stands for the expected values given by the prediction of canonical ensemble. In both plots (a) and (b), we mark the scar states in the red circles. Plots (c) and (d) demonstrate the deviation between the eigenstates expectations of local observables $\hat{O}_1,\hat{O}_2$ and that of the DMs: $\text{Tr}(\hat{O}_{1,2}\rho)-\langle \hat{O}_{1,2} \rangle_i$. We take $\rho$ to be either the canonical DM or the adjusted version from our work. The results from the canonical DM $\rho_c(\beta_c)$ is represented by the gray and black circles, with the former marking the thermal states, and the later marking the scar states. The results from the adjusted grand canonical DM $\rho(\beta,\nu)$ are represented by blue dots, where the scar states are highlighted with red circles. We choose $\hat{O}_1$ to be the single site spin-$z$ operator $\hat{O}_1=s_k^z$, and $\hat{O}_2$ to be the product of spin-$z$ operators on the next nearing sites $\hat{O}_2=s_k^z\otimes s_{k+2}^z$. We carried out our calculations in a spin-$1$ chain with 12 sites.}
	\label{fig: hPXP}
\end{figure}

\section{Application in other models hosting QMBS}
{\bf spin-$3/2$ scenario.} The calculations from Fig.~4 can be generalized to the corresponding model featuring larger spins. In Fig.~\ref{fig: spin15expectation_and_schatten_distance},  we provide the results calculated for the aforementioned model in the spin-$3/2$ scenario. The calculation shows that the grand canonical ensembles fit the local properties of the eigenstates better than the canonical ensemble, confirming our discussion in section IV.

{\bf High-dimensional PXP model.} Our framework can be applied to other constrained systems. Here we consider another QMBS model based on 1D spin chain, which is known as high-dimensional PXP model. The Hamiltonian writes:
\be\label{origHhpxp}
\bal
      H_{hPXP} & =\hat{P} \sum_{k=1}^{N}s_k^x \hat{P}, \\
      \hat{P} & =\prod_{k=1}^{N}(\mathbb{I}-T_k\otimes T_{k+1}), \\
      T_k & =\sum_{m=-j+1}^{j}\left| m \right\rangle\left\langle m \right|_k,
\eal
\ee
where $\hat{P}$ stands for the projector of the constrained subspace. $\mathcal{B}=\{ \pi_k=T_k\otimes T_{k+1}, k=1,2,\cdots, N \}$ mark the blockades that prohibit nearest neighbouring spins from spontaneously occupying states $\{ |-j+1\rangle_k,|-j+2\rangle_k,\cdots,|j\rangle_k \}$. $|m\rangle_k$ denotes the eigenstate of $s_k^z$: $s_k^z|m\rangle_k=m|m\rangle_k$. Therefore a spin remains dynamically constrained unless its two neighboring spins are both in the lowest spin-$z$ eigenstates $|-j\rangle_k$.

Our framework can be applied to this model directly.
However, solving the energy spectrum of this model may pose significant challenges. Considering a spin-$j$ chain with $N$ sites, the dimension of the corresponding Hilbert space is $(2j+1)^N$ with numerous unnecessary degrees of freedom. 
We can reduce these redundant degrees of freedom by re-encoding two adjacent spins $|m,m'\rangle_{2l-1,2l}$ into a new logic site $|\cdot\rangle_l$ with $4j+1$ degree of freedom. 
Formally, we can divide the blockades $\mathcal{B}$ into two subsets of blockade projectors: $\mathcal{B}_{o}=\{\pi_{2l-1},l=1,2,\cdots,N/2\}$ and $\mathcal{B}_{e}=\{\pi_{2l},l=1,2,\cdots,N/2\}$, where $\hat{B}_{o}$($\hat{B}_{e}$) act only on the neighboring $2l-1$-th and $2l$-th sites($2l$-th and $2l+1$-th sites).
The projector of the constrained subspace can also be decomposed into two parts: $\hat{P}=\hat{P}_{o}\cdot\hat{P}_{e}$, where $\hat{P}_{o(e)}=\prod_{\pi_l\in \mathcal{B}_{o(e)}}(\mathbb{I}-\pi_l)$. Following this, we can rewrite the Hamiltonian as:
\be\label{decomposeHhpxp}
\bal
  & H_{hPXP}=\hat{P}_e \hat{P}_o \sum_{k=1}^{N}s_k^x \hat{P}_o \hat{P}_e = \hat{P}_e H_{odd} \hat{P}_e, \\
  & H_{odd}=\hat{P}_o \sum_{k=1}^{N}s_k^x \hat{P}_o=H_0\cdot \hat{P}_o=\hat{P}_o\cdot H_0,
\eal
\ee
where $H_0$ commutes with $\hat{P}_o$: $[H_0,\hat{P}_o]=0$, and is equivalent to $H_{odd}$ within the subspace denoted by $\hat{P}_{o}$. Considering the null space of $\hat{P}_o$, $H_{odd}$ is entirely constrained while $H_0$ can still drive the system evolving in the unconstrained area. Moreover, $H_0=\sum_{l=1}^{N/2}h_l$ is comprised of local terms $h_l$'s, which signify the blockade $\mathcal{B}_o$ with local interactions between neighboring spins:
\be\label{rehpxp}
\bal
 h_l & = (\mathbb{I}-\pi_{2l-1})(s_{2l-1}^x + s_{2l}^x)(\mathbb{I}-\pi_{2l-1}) \\
     & =(\mathbb{I}-T_{2l-1}\otimes T_{2l})(s_{2l-1}^x + s_{2l}^x)(\mathbb{I}-T_{2l-1}\otimes T_{2l}) \\
     & =(\mathbb{I}-T_{2l-1})\otimes s_{2l}^{x}+s_{2l-1}^{x}\otimes (\mathbb{I}-T_{2l}) \\
     & = |-j\rangle\langle -j|_{2l-1}\otimes s_{2l}^x+s_{2l-1}^x \otimes|-j\rangle\langle -j|_{2l}.
\eal
\ee
Then we rewrite the constrained Hamiltonian as:
\be\label{rehpxpHH}
\bal
  H_{hPXP} & = \hat{P}_e H_{odd} \hat{P}_e = \hat{P}_e H_0 \hat{P}_e\cdot \hat{P}_o \\
           & = \sum_{l=1}^{N/2} \prod_{l_1,l_2=l-1}^{l}(\mathbb{I}-\pi_{2l_1})h_l(\mathbb{I}-\pi_{2l_2}) \cdot \hat{P}\\
           & = \sum_{l=1}^{N/2} \prod_{l_1,l_2=l-1}^{l}(\hat{P}_o-\pi_{2l_1}\cdot \hat{P}_o)h_l(\hat{P}_o-\pi_{2l_2}\cdot \hat{P}_o) \cdot \hat{P}\\
           & = \sum_{l=1}^{N/2} \prod_{l_1,l_2=l-1}^{l}(\mathbb{I}-\pi'_{2l_1})h_l(\mathbb{I}-\pi'_{2l_2}) \cdot \hat{P} \\
           & = H\cdot \hat{P}, \\
  H        & = \sum_{l=1}^{N/2}h_l-\Pi_l h_l - h_l \Pi_l + \Pi_l h_l \Pi_l,
\eal
\ee
where we have utilized the commutation relation between $\hat{P}_o$ and $\hat{P}_e$ in the first line. Besides, $\hat{P}_o^2=\hat{P}_o$ is a projector, which enables us to project the blockades $\pi_{2l}$'s onto the subspace denoted by $\hat{P}_o$ and arrive at $\pi'=\pi \cdot \hat{P}_o$. This further leads us to define the local blockade projectors $\Pi_l=\pi'_{2l-2}+\pi'_{2l}$. Finally, $H$ commutes with $\hat{P}$ and is the equivalent Hamiltonian for this model.

Now we take the critical step by viewing the subspace denoted by $\hat{P}_o$ as the entire Hilbert space. Consequently, we can regard $h_l$ as the local Hamiltonian acting on the $l$-th logic site, and consider this model as an outcome of applying the blockades $\mathcal{B}_e$ on $H_0$. Correspondingly, we should eliminate the states $|m,m'\rangle_{2l-1,2l}$ with $m$ and $m'$ both greater than $-j$, since these states are located in the null space of $\hat{P}_o$. Then we map the rest $4j+1$ states as:
\begin{equation}\label{hPXPlogicsite}
 \begin{aligned}
 & \left| -j,m \right\rangle_{2l-1,2l}\rightarrow\left| a_m \right\rangle_{l}, \ m=-j+1,\cdots,j-1,j ,\\
 & \left| m,-j \right\rangle_{2l-1,2l}\rightarrow\left| b_m \right\rangle_{l}, \ m=-j+1,\cdots,j-1,j ,\\
 & \left| -j,-j \right\rangle_{2l-1,2l}\rightarrow\left| c \right\rangle_{l},
 \end{aligned}
\end{equation}
with $l=1,2,\cdots N/2$ denotes the logic site number. 
We note that all the terms in $H$ can be expressed with these states. Therefore, we have effectively reduced the dimension of the Hilbert space to $(4j+1)^{N/2}$.

Following the framework and Eq.(\ref{rehpxpHH}), we can build up a master equation equivalent to the unitary dynamics governed by $H$, and construct the quasi-particle counting operator $\hat{N}=\sum_{k=1}^{N/2}L_{k,1}^{\dagger}L_{k,1}$, which then leads to the alternation of the Gibbs ensemble as discussed in section V.
We spare the repetitive derivation process and provide the calculations done upon this model below.

First, we show the eigenstates' deviation from ETH by calculating the expectation of the local observable $N_k=L_{k,1}^{\dagger}L_{k,1}$ in Fig.~\ref{fig: hPXP}(b). We also show the overlap between the eigenstates and the appointed initial state in Fig.~\ref{fig: hPXP}(a) as a reference. Scar eigenstates appear to be the ones that show higher overlaps with the initial state, and at the same time, mostly diverge from the canonical average.
Then we display the deviation between the eigenstates' expectation values and the ensemble average for both the grand canonical ensemble and the canonical ensemble in Fig.~\ref{fig: hPXP}(c) and (d). We have employed two local observables for the calculation. It can be seen that the grand canonical ensemble shows significant advantage over the canonical ensemble, as the grand canonical DM can accurately predict the expectation values for both the scar states and the thermal eigenstates. Specifically, the grand canonical ensemble provided in Eq.(13) demonstrates better fits for the local properties of the scar eigenstates, while these properties show significant deviation from the canonical ensemble average originated from ETH.
